\documentclass[prl,amsmath,a4paper,twocolumn]{revtex4}
\usepackage{graphicx}
\usepackage{amsfonts}
\usepackage{amsmath}
\usepackage{ltxtable}
\usepackage{CJK}
\usepackage{mdwmath}
\usepackage{color}

\newcommand{\mB}{{\mathcal B}}

\begin{document}

\title{Greenberger-Horne-Zeilinger Paradoxes from Qudit Graph States }
\author{Weidong Tang$^{1,2}$, Sixia Yu$^{1,2}$ and C.H. Oh$^{2}$}
\affiliation{$^1$Hefei National Laboratory for Physical Sciences at
Microscale and Department of Modern Physics
of University of Science and Technology of China, Hefei 230026, P.R. China\\
$^2$Centre for Quantum Technologies and Physics Department,
National University of Singapore, 2 Science Drive 3, Singapore 117542}
\begin{abstract}
One fascinating way of revealing  quantum nonlocality is the all-versus-nothing test due to Greenberger, Horne, and Zeilinger (GHZ) known as GHZ paradox. So far genuine multipartite and multilevel GHZ paradoxes are known to exist only in systems containing an odd number of particles. Here we shall construct GHZ paradoxes for an arbitrary number (greater than 3) of particles with the help of qudit graph states on a special kind of graphs, called  GHZ graphs.  Furthermore, based on the GHZ paradox arising from a GHZ graph, we derive a Bell inequality with two $d$-outcome observables for each observer, whose maximal violation attained by the corresponding graph state, and a Kochen-Specker inequality testing the quantum contextuality in a state-independent fashion.

\end{abstract}

\maketitle
Local realism cannot make quantum theory complete, as argued by Einstein, Podolsky, and Rosen (EPR) based on the belief that every element of physical reality must have a counterpart in  a complete theory \cite{EPR}. According to them, an element of reality is corresponding to a physical quantity whose value can be predicted with certainty without in any way disturbing a system. No disturbance is ensured by the locality, i.e., the assumption that the result of a measurement cannot be affected by any spacelike separated events. The clashing between the local realism  and quantum mechanics as revealed by several no-go theorems such as Bell's theorem \cite{Bell},  Greenberger-Horne-Zeilinger (GHZ) theorem \cite{GHZ,GHZ2,mermin}, and Kochen-Specker (KS) theorem \cite{KS}, shows that the quantum mechanical description of our world is nonlocal, or more generally contextual. This fascinating and fundamental quantum feature of nonlocality and contextuality has been verified in experiments on various physical systems, e.g., \cite{bell exp}, via the detection of violations of Bell inequalities and KS inequalities  \cite{cabello2,pit,yu-oh}.

Among these genius approaches, GHZ theorem  \cite{GHZ,GHZ2} provides us an ``all-versus-nothing" \cite{AVN} test of a stronger type nonlocality, referred to as GHZ nonlocality, than Bell's nonlocality. This is a state-dependent argument: because of the perfect correlations in some special state called the GHZ state, e.g., a 3-qubit GHZ state  $|\Phi\rangle=\frac{1}{\sqrt{2}}(|000\rangle-|111\rangle)$, some local observables are elements of reality according to EPR. For example since the observable $\sigma_x^1\sigma_y^2\sigma_y^3$ stabilizes the GHZ state, i.e., $\sigma_x^1\sigma_y^2\sigma_y^3|\Phi\rangle=|\Phi\rangle$,  observables $\sigma_x^1, \sigma_y^2, \sigma_y^3$  are all elements of reality. Here $\sigma_{x,y,z}^k$ denote 3 standard Pauli matrices for the $k$th qubit. Similarly, from two other stabilizers  $\sigma_y^1\sigma_x^2\sigma_y^3$ and $\sigma_y^1\sigma_y^2\sigma_x^3$ of $|\Phi\rangle$ we know that all $\sigma_{x,y}^k$ are elements of reality and must have realistic values $m^{x,y}_k=\pm1$ for $k=1,2,3$. Realistic values are supposed to obey the same algebraic relations as their corresponding observables. That is to say we have on the one hand $m^x_1m^x_2m^x_3=-1$, since $\sigma_x^1\sigma_x^2\sigma_x^3|\Phi\rangle=-|\Phi\rangle$ and $m_1^xm_2^ym_3^y=$$m_1^ym_2^xm_3^y=$$m_1^ym_2^ym_3^x=1$. On the other hand, since $(m_k^{y})^2=1$, we have identity $m^x_1m^x_2m^x_3=(m^x_1m^y_2m^y_3)(m^y_1m^x_2m_3^y)(m_1^ym_2^ym_3^x)$ which gives rise to a contradiction $-1=1$.

This elegant presentation of the GHZ paradox for 3 qubits is due to Mermin \cite{mermin} soon after its first discovery for a 4-qubit GHZ state \cite{GHZ} and has already been verified experimentally \cite{GHZ exp}.
 Although originally the GHZ argument is state dependent, it was found recently that any GHZ paradox can give rise to a KS inequality for a state-independent test of quantum contextuality \cite{cabello2}. In addition to its fundamental role played in our understanding of quantum nonlocality and contexuality, the GHZ paradox also finds numerous applications such as in the quantum protocols for reducing communication complexity \cite{cleve} and for secret sharing \cite{secret}.

Compared to the bipartite and two-level case,
multipartite and multilevel nonlocality or entanglement is poorly
understood. In some quantum informational tasks such as quantum
cryptography, the usage of multidimensional systems offers
advantages such as an increased level of tolerance to noise at a given
level of security and a higher flux of information compared to the two dimensional case\cite{simon}.
Thus, it is crucial to investigate the relevant physical properties from some
subclasses of these systems, e.g., GHZ nonlocality from a special kind of
qudit states. Earlier efforts \cite{cabello,pagonis} to generalize GHZ paradoxes to multidimensional and multilevel systems can be reduced either to the qubit cases or to fewer particle cases, except the cases of $n=4j+3$ for qubits \cite{pagonis}. Genuine multipartite multilevel GHZ paradoxes were first found by Cerf {\it et al}. for $(d+1)$-partite $d$-level systems with $d$ being even \cite{cerf}. An unconventional approach by using concurrent observables, not commuting yet having a common eigenstate, is proposed by Lee {\it et al.} to construct a GHZ paradox for the GHZ states of an odd number of particles \cite{lee}. Also a GHZ-like argument (all-versus-something) is proposed by Kaszlikowski {\it et al.} for $d$-partite $d$-level systems \cite{NN}, in which concurrent observables have been used implicitly. Later, DiVincenzo and Peres \cite{divi} found out that not only can GHZ states exhibit the GHZ paradox but also those code words, which are one kind of multipartite entangled
states used in quantum error corrections \cite{QECC}, can exhibit GHZ nonlocality. But so far genuine multipartite and multilevel GHZ paradoxes for an even number of particles are still missing.

It turns out that GHZ states as well as code words from stabilizer codes \cite{stab} are graph states \cite{Briegel} which are essential resources for the one-way computing \cite{one way} and also provide an efficient construction of quantum error-correcting codes \cite{yu0}. It is thus natural to take advantage of the perfect correlations in graph states for the constructions of GHZ paradoxes.
 In this Letter we shall identify those graphs, called  GHZ graphs, whose corresponding graph states lead to genuine multipartite multilevel GHZ paradoxes. Furthermore we derive a Bell inequality for multipartite and  multilevel systems as well as a state-independent KS inequality for every GHZ graph.

As a graph state for qubits is related to a simple graph, a nonbinary graph state \cite{werner,schli,Hu} is associated with a weighted graph.
Let $\mathbb Z_d=\{0,1,\ldots, d-1\}$ denote the ring with addition modulo $d$. A {\it $\mathbb Z_d$-weighted} graph $G=(V,\Gamma)$ is composed of a set $V$ of $n$ vertices and a set of weighted edges specified by the
adjacency matrix $\Gamma$, a symmetric $n\times n$  matrix with zero diagonal entries and the matrix element
$\Gamma_{uv}\in\mathbb Z_d$ denoting the weight of the edge connecting the vertices $u$ and $v$.
A graph is {\it connected} if for any pair of vertices $u,v$ there exists a finite number of vertices $\{v_i\}_{i=0}^K$ such that $\prod_{i=0}^{K-1}\Gamma_{v_iv_{i+1}}\not=0$ with $u=v_0$ and $v=v_K$.

We denote by $D_v$ the degree of vertex $v\in V$ which is the sum of the weights of all the edges connecting to $v$ and by $W$ the {\it total weight} of $G$ which is the sum of the weights of all the edges. Explicitly, we have
\begin{eqnarray}
D_v=\sum_{u\in V}\Gamma_{uv}\ (v\in V), \quad W=\frac12\sum_{u,v\in V}\Gamma_{uv}.
\end{eqnarray}
A  GHZ graph is a connected $\mathbb Z_d$-weighted graph satisfying {\it (i) the degree  of each vertex is divisible by $d$, i.e., $D_v\equiv 0\mod d$, while (ii) the total weight is NOT divisible by $d$; i.e., $W\not\equiv 0\mod d$}. From these two conditions it follows immediately that the GHZ graph does not exist in odd dimensions and $\omega ^{W}=-1$, where $\omega=e^{ i\frac {2\pi}d}$. In fact, from the first condition, there is an integer $t_v$ such that $D_v=dt_v$ for each $v\in V$, and from the fact that the total weight $W=dt/2$ with $t=\sum_{v\in V}t_v$ is an integer, since $\Gamma$ is symmetric, it follows that if $d$ is odd then $t$ must be even  and thus $W$ is divisible by $d$. Furthermore, in even dimensions, the total weight $W$ is not divisible by $d$ if and only if $t$ is odd and thus $\omega^W=(-1)^t=-1$.
In what follows we shall always assume $d$ to be even. A GHZ graph is called  `` primary" if for each vertex $a\in V$ there exists a pair of vertices $b,c$ such that $\Gamma_{ab}$ and $\Gamma_{ac}$ are coprime and ``weakly primary"  if there exist three vertices $a,b,c\in V$, such that $\Gamma_{ab}$ is coprime with $\Gamma_{ac}$.

\begin{figure}
\includegraphics[scale=0.8]{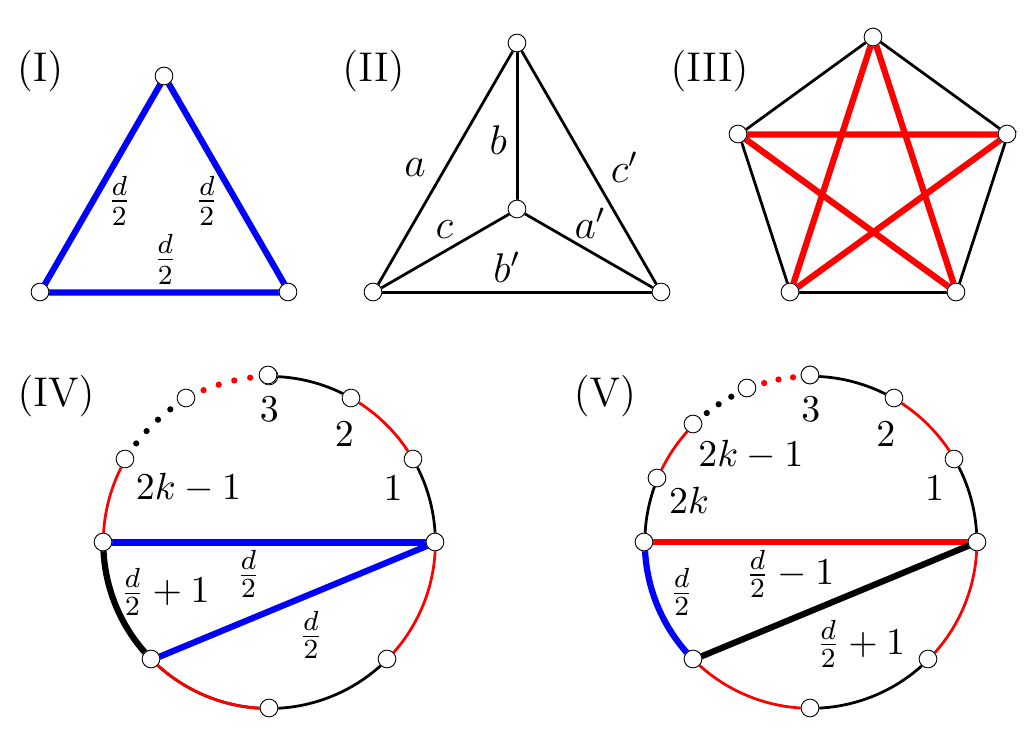} \caption{\label{GHZgraph} Examples of GHZ graphs.  Unlabeled thin black or red edges have weight $1$ or $d-1$, respectively. All possible GHZ graphs on 3 and 4 vertices are shown in (I) and (II), where $a^\prime=\frac d2+a$, $b^\prime=\frac d2+b$, and $c^\prime=\frac d2+c$ with $a+b+c=d/2$. A GHZ graph on 5 vertices is shown in (III) where the thick red edges have weight $d/2-1$. In (IV) and (V) two primary GHZ graphs on $2k+4$ and $2k+5$ vertices $(k\ge 1)$ are shown.}
\end{figure}

In the case of $d=2$ a GHZ graph has an odd number of edges and every vertex has an even number of neighbors. All GHZ graphs for $d=2$ are primary. For example, a loop graph with an odd number of vertices and a complete graph with $4j+3$ ($j\ge 0$) vertices are possible GHZ graphs. There is only a single  GHZ graph on 3 vertices as shown in Fig.1(I) and it is clear that it is not weakly primary if $d>2$. In the case of $n=4$ all possible GHZ graphs are shown in Fig.1(II) with weights satisfying $a+b+c=d/2$. If $d=4k$ then $d/2\pm 1=2k\pm1$ are coprime and thus, by choosing, e.g., $a=1,c=1$, we obtain a primary GHZ graph. If $d=4k+2$ then there always exists a vertex with all edges having even weights, since $d/2$ is odd, so that only a weakly primary GHZ graph exists in this case. Examples of primary GHZ graphs for arbitrary $n\ge 5$ and even dimensions are shown in Fig.1(III-V). The primary GHZ graph on 5 vertices as shown in Fig.1(III) can be generalized to any odd number of vertices.

Consider a system of $n$ particles each of which has $d$ energy levels, a {\it qudit} for short, and label them with $V$. Let $\{|s\rangle_v|s\in \mathbb Z_d\}$ be the computational basis for qudit $v\in V$ and $\{|{\bf s}\rangle|{\bf s}\in \mathbb Z_d^V\}$ is a basis for $n$ qudits where $\mathbb Z_d^V$ is the set of all $n$-dimensional vectors ${\bf s}=(s_1,s_2,\ldots,s_n)$  with components $s_v\in\mathbb Z_d$ for all $v\in V$. To any weighted graph $G=(V,\Gamma)$ on $|V|=n$ vertices we can associate with a qudit graph state
\begin{equation}
|\Gamma\rangle=\frac1{d^{\frac n2}}\sum_{{\bf s}\in \mathbb Z^V_d}\omega^{\frac12{\textbf{s}\cdot\Gamma\cdot\textbf{s}}}|{\bf s}\rangle,
\end{equation}
which is also the unique joint $+1$ eigenstate of $n$ commuting vertex stabilizers
\begin{equation}
g_v=X_v\prod_{u\in V}Z_u^{\Gamma_{uv}},
\end{equation}
i.e, $g_v|\Gamma\rangle=|\Gamma\rangle$  for all $v\in V$. Here
we have introduced the generalized
bit shift operator $
X_v=\sum_{s\in\mathbb Z_d}|(s+1)\mod d\rangle\langle s|_v$ and phase shift operators
$
Z_v=\sum_{s\in\mathbb Z_d}\omega^{s}|s\rangle\langle
s|_v$ for each qudit $v\in V$.
It is easy to check that $X^d_v=Z^d_v=I$ and $Z_vX_v=\omega{X_vZ_v}$. Our main result reads as follows:

{\it Theorem }  For each (weakly) primary GHZ graph $G=(V,\Gamma)$ on $|V|=n$ vertices, with weights taken values in $\mathbb Z_d$, the graph state $|\Gamma\rangle$ provides a (weakly) genuine $n$-partite $d$-level GHZ paradox.

Before embarking on the proof we should clarify what we mean by {\it genuine $n$-partite and $d$-level} and give an example. According to \cite{cerf} a GHZ paradox, formulated via a set of commuting observables, is said to be genuinely $n$-partite if one cannot reduce the number of parties and still have a Mermin-GHZ paradox. A GHZ paradox is (weakly) genuine $d$-level if one cannot reduce the dimensionality of the Hilbert space of (all) any one of the parties to less than $d$ and still have a paradox.

As an example let us consider the GHZ graph as shown in Fig.1(II) in the case of $n=4$ and the following 5 commuting observables that stabilize the corresponding graph state
\begin{equation}\label{GHZ-4}
\begin{array}{ccccc}
X&Z^{a^\prime}&Z^b&Z^c&+1\cr
Z^{a^\prime}&X&Z^{c^\prime}&Z^{b^\prime}&+1\cr
Z^{b}&Z^{c^\prime}&X&Z^{a}&+1\cr
Z^c&Z^{b^\prime}&Z^{a}&X&+1\cr
X^\dagger&X^\dagger&X^\dagger&X^\dagger&-1
\end{array}
\end{equation}
which provide us a GHZ paradox. Measurement of the product of the operators in each row gives a certainty result $1$ or $-1$ as listed in the right column of Eq.(\ref{GHZ-4}) by quantum mechanics. With the analogue to EPR's argument, the result $m_v^x$ or $m_v^z$ of measuring the corresponding $d$-outcomes measurements $X_v$ or $Z_v$ on the $v$th qudit  can be predicted in advance with certainty with the help of  the results of spacelike separated measurements of $X$ or $Z$ on the other three qudits and are therefore elements of reality. Because the algebraic relations are preserved, we have $m_v^{x,z}=\omega^k$ with $\omega=e^{i2\pi/d}$ for some $k\in \mathbb Z_d$ and
\begin{align}\label{assignV}
(m^{x}_1) (m^{z}_2)^{a^\prime} (m^{z}_3)^b (m^{z}_4)^c&=1\cr
(m^{z}_1)^{a^\prime} (m^{x}_2) (m^{z}_3)^{c^\prime} (m^{z}_4)^{b^\prime}&=1\cr
(m^{z}_1)^{b} (m^{z}_2)^{c^\prime} (m^{x}_3) (m^{z}_4)^{a}&=1\cr
(m^{z}_1)^c (m^{z}_2)^{b^\prime} (m^{z}_3)^{a} (m^{x}_4)&=1\cr
(m^{x}_1)^{-1} (m^{x}_2)^{-1} (m^{x}_3)^{-1} (m^{x}_4)^{-1}&=-1.
\end{align}
The contradiction lies in the fact that all five  equations in Eq.(\ref{assignV}) cannot hold simultaneously.
In the case of $d=4$ if we choose $a=b=1$ and $c^\prime=2$ with $a^\prime=b^\prime=3$ and $c=0$ then the GHZ graph is primary and the corresponding GHZ paradox is genuine 4-partite and 4-level.
In the case of $d=6$ we can choose $a=b=c=1$ and $a^\prime=b^\prime=c^\prime=4$ such that for the second qudit there exists a projection to a qutrit by identification $Z^2$ for $d=6$ with $Z$ for $d=3$. Thus it provides an example of weakly genuine 6-level GHZ paradox that can be regarded as GHZ paradox on a hybrid system of three 6-level system plus a qutrit.

{\it Proof}.--- Let $G=(V, \Gamma)$ be a GHZ graph; i.e., the degree of each vertex $D_a$ is divisible by $d$ and  the total weight $W$ satisfies $\omega^{W}=-1$. For each qudit $v\in V$ we measure two unitary observables $X_v$ and $Z_v$ with outcomes assigned to values $m^{x}_v,m^{z}_v\in\{\omega^t|t\in \mathbb Z_d\}$, respectively. First of all these values are elements of reality because of the perfect correlations $g_v|\Gamma\rangle=|\Gamma\rangle$ $(v\in V)$. In any local, or noncontextual, hidden variable models these values are independent of which observables might be measured by other observers. Furthermore, they must satisfy the same algebraic rules, e.g., the product rule, as their quantum counterparts do. For example from the definition of the vertex stabilizer $g_v$  it follows
\begin{equation}
M_v:=m_v^x\prod_{u\in V}(m_u^z)^{\Gamma_{uv}}=1
 \end{equation}
for each $v\in V$. On the other hand from the constraint
$X_V|\Gamma\rangle=-|\Gamma\rangle$,
 because of  the identity
\begin{equation}
\prod_{a\in V}g_{a}
=\omega^{W}X_{V}\prod_{a\in V}Z_a^{D_a}=-X_V,
\end{equation}
it follows that $\prod_{v\in V}m_v^x=-1$ which is impossible because $\prod_{v\in V}M_v=\prod_{v\in V}m_v^x$, in which the fact that $D_v$ is divisible by $d$ has been used.

By definition a GHZ graph is a connected graph and thus for each partition of $n$ observers into two groups some of $n+1$ unitary observables will not be commuting when restricting to either one of two groups. Therefore  the GHZ paradox for $|\Gamma\rangle$  is a genuine $n$ partite. Furthermore, if the GHZ graph is primary then each vertex is attached to at least one pair of edges of coprime weights. If there were a projection to lower dimensions for a qudit,  some eigenstates of $X_a$ and those of $Z_a^{\Gamma_{ab}}$ and $Z_a^{\Gamma_{ac}}$ are orthogonal. This is impossible because first there always exist $p,q\in \mathbb Z_d$ such that $p\Gamma_{ab}+q\Gamma_{ac}=1\mod d$ and, second, $X_v$ and $Z_v$ are two complementary observables whose eigenstates cannot have a zero overlap, which is the case if the dimensionality can be reduced. \hfill$\sharp$

Some remarks are in order. First, for  we have constructed genuine $n$-partite and $d$-level GHZ paradox with $n\ge 5$ can be even. Second, any state that is related with GHZ graph states via local unitary transformations exhibits also GHZ nonlocality. Third,
for a graph that is not GHZ graph it is also possible to construct a GHZ paradox for the graph state if the underlying graph contains a GHZ subgraph.
A \emph{subgraph} $H=(V^\prime,\Gamma^\prime)$ of a weighted graph $G=(V,\Gamma)$ is also a $\mathbb Z_d$-weighted graph with a vertex set given by $V^\prime\subseteq V$ and edges specified by $\Gamma^\prime_{ab}=\Gamma_{ab}$ if $a,b\in V^\prime.$ If furthermore the subgraph is a GHZ graph we shall refer to it as a {\it GHZ subgraph} of $G$. Suppose that the graph $G$ contains a GHZ graph $H=(V^\prime,\Gamma^\prime)$ with $|V^\prime|=m<n$, then the $m+1$ observables $g_u$ with $u\in V^\prime$ and $\prod_{u\in V^\prime}g_u$ yield a GHZ paradox for the graph state $|\Gamma\rangle$. It is clear that it is only a genuine $m$-partite GHZ paradox if the GHZ subgraph is primary. For example, the 4-qubit GHZ state is equivalent to the graph state corresponding to  the complete graph on 4 vertices, which contains a loop of length 3 as a GHZ subgraph. In fact the original GHZ proof \cite{GHZ2} revealed a 3-partite GHZ nonlocality using this GHZ subgraph.

As the first application we shall derive a Bell inequality with two measurement settings for each observer with the help of the GHZ paradox derived from a GHZ graph. Consider two $d$-outcome measurements $A_v$ and $B_v$ for each observer $v\in V$ and assign values in $\{\omega^t|t\in \mathbb Z_d\}$ to them (Bell-KS value assignment). For each GHZ graph $G=(V,\Gamma)$ we introduce a Bell operator as

 \begin{eqnarray}
\mB_G=\mathop{\sum_{k=1}^{d-1}}_{k\ odd}\frac{2}d\left(
\sum_{v\in V}A_v^k\prod_{u\in V}B_{u}^{k\Gamma_{uv}}-\prod_{v\in V} A_v^k\right)
\end{eqnarray}
Taking into account the identity $\sum_{k=0}^{d-1}\omega^{k l}=d\delta_{l,0}$ for arbitrary $l\in \mathbb Z_d$ and denoting $A_V=\prod_{v\in V}A_v$ and $N_v=A_v\prod_{u\in V}B_u^{\Gamma_{uv}}$ for each $v\in V$, where $\delta_{i,j}$ is the standard Kronecker delta symbol,  we have
\begin{equation}
\mB_G=\delta_{-1,A_V}-\delta_{1,A_V}+\sum_{v\in V}\left(\delta_{1,N_v}-
\delta_{-1,N_v}\right)\le n-1.
\end{equation}
The inequality holds in any local realistic theory because if there are $n$ positive terms then there is necessarily a negative term in $\mB_G$: If $N_v=1$ for every $v\in V$ then it holds $A_V=1$ which contributes a negative term; if $N_v=1$ for all $v\in V-\{v_0\}$ and $A_V=-1$ then it necessarily holds $N_{v_0}=-1$ because $A_V=\prod_{v\in V}N_v$. Furthermore it is easy to see that $\mB_G\le n+1$, which is attained by the graph state $|\Gamma\rangle$ with $\langle\Gamma| \mB_G|\Gamma\rangle=n+1$ in which $A_v$ and $B_v$ are chosen to be $X_v$ and $Z_v$, respectively, for each $v\in V$. In this case the quantum to classical ratio $(n+1)/(n-1)$ is a constant independent of the dimension, comparing to that of \cite{son}.

Every GHZ paradox leads also to a proof of KS theorem. And any proof of KS theorem can be converted to an experimentally testable inequality, called as KS inequality, in the manner of Cabello \cite{cabello2}. As the second application we consider the following KS inequality
\begin{eqnarray}\label{ksub}
\frac12\left\langle X_V^\dagger \prod_{v\in V}X_{v}+\sum_{v\in V}g_v^\dagger X_v \prod_{u\in V}Z_{u}^{\Gamma_{uv}}+ H.c.\right\rangle_{c}\cr-\frac12\left\langle X_V^\dagger \prod_{v\in V} g_v+H.c.\right\rangle_{c}\le C_{n+1,d}
\end{eqnarray}
where,  with $\lambda=\frac{d}{2(n+1)}$ and $\theta=2\pi/d$, we have denoted
\begin{equation}
\frac{C_{nd}}{n+1}=(\lambda-\lfloor\lambda\rfloor)\cos \lceil\lambda\rceil\theta+
\left(1+\lfloor\lambda\rfloor-\lambda\right)\cos \lfloor\lambda\rfloor\theta.
\end{equation}
First, each term, e.g.,$\left\langle X_V^\dagger \prod_{v\in V} g_v\right\rangle_{c}$, is the abbreviated form of the classical correlation of $n+1$ observables, e.g., $\left\langle X_V^\dagger g_1 g_2 ... g_n\right\rangle_{c}$.  Second, the upper bound can be easily inferred from the Lemma proved below. Third, we have $C_{n+1,d}<n+2$ while the quantum mechanical value of the left-hand side of Eq.(\ref{ksub}) equals to $n+2$ identically and therefore violates the above KS inequality in a state-independent fashion.

In summary, first of all we have identified a special kind of graphs, called  GHZ graphs, whose corresponding graph states give rise to GHZ paradoxes. Except for the case $n=4$ with $d=4k+2$ for which only a weakly genuine GHZ paradox is found we have derived genuine $n$-partite and $d$-level GHZ paradoxes from qudit graph states corresponding to GHZ graphs with $n\ge 4$ and even $d$ being arbitrary. Second, as applications
for each GHZ graph we derive a Bell inequality with two $d$-outcome observables for each observer whose maximal violation is attained by the corresponding graph state as well as a state-independent KS inequality that is satisfied by any noncontextual hidden variable models. This would be helpful to the analysis of multipartite contextuality or multipartite nonlocality. It should be noted that GHZ paradoxes may exist for those states that are equivalent to the graph states under local Clifford (LC) transformations. However the conditions under which both two GHZ paradoxes arising from two LC equivalent states are genuine $n$-partite  seem to lie out of the reach of current Letter.
Besides, the examples we are analyzing here involve only some special classes of graph states, so figuring out other classes of graph states which are consistent with our theorem are still meaningful as for  fixed parties $n$, different graphs may have different robustness against decoherence, which may help to design new quantum protocols for reducing communication complexity. Ironically, a genuine 4-partite GHZ paradox is still missing for the original 4-qubit GHZ state.

This work is supported by National
Research Foundation and Ministry of Education, Singapore
(Grant No. WBS: R-710-000-008-271) and supported by the financial support of NNSF of China (Grant No. 11075227).

\newpage
 {\it Lemma } Let $\mathbb U_d=\{\theta,2\theta,\ldots,d\theta\}$ with $\theta=2\pi/d$  and $\lambda=\frac d{2(n+1)}$ and $\vec x=(x_1,x_2,\ldots,x_n)\in \mathbb R^n$  be $n$ real variables and $x_s=\sum_{i=1}^nx_i$. We have
\begin{eqnarray}
\max_{\vec x\in \mathbb U^n_d}\left\{f(\vec x):=\sum_{i=1}^n \cos x_i-\cos\left(x_s\right)\right\}=C_{n,d}.
\end{eqnarray}
Specially, if $n\ge d/2$, i.e., $\lambda<1$, then $C_{nd}=n+1-d\sin^2\frac\pi d$ and if $d=2(n+1)l $ for some $l$, i.e., $\lambda$ is an integer, then $C_{nd}=(n+1)\cos\frac\pi{n+1}$.

{\it Proof. } The maximum of $f(\vec x)$ over $\mathbb R^n$ is the largest value on all
local extremal points satisfying
$\frac{\partial f(\vec x)}{\partial x_i}=\sin x_s-\sin x_i=0\ (\forall i).$
Let $x_1=x$ then either $x_i=a_i=2u_i\pi+x$ or $x_i=b_i=(2u_i+1)\pi-x$  for all integers $u_i$ with $i\ge 2$ since $\sin x_i=\sin x$. Denote by $m$ the number of $x_i$'s being equal to $a_i$ and $k=n-m$ the number of $x_i$'s being equal to $b_i$'s among $\{x_i\}_{i=1}^n$. Then from $\sin x=\sin (k\pi+(m-k)x)$ it follows either a) $x=x_a$ with $2l\pi+x_a=k\pi+(m-k)x_a$ or b) $x=x_{b}$ with  $(2l+1)\pi-x_b=k\pi+(m-k)x_b$ for all the integers $l\ge0$. At these extremal points we have either $f(\vec x_a)=(m-k-1)\cos x_a$ or $f(\vec x_b)=(m-k+1)\cos x_{b}$. If $k\ge 1$ then $m-k=n-2k\le n-2$ and thus $f(\vec x_{a,b})\le n-1$. If $k=0$ then
$
f(\vec x_l)=(n+1)\cos x_l,
$
where $\vec x_l=(x_l,x_l,\ldots,x_l)$ with $x_l=(2l+1)\pi/(n+1)$. Since $f(\vec x_0)\ge n-1$ and $f(\vec x_0)\ge f(\vec x_l)$ the extremal point $\vec x_0$ leads to the largest value of $f$.

If $\lambda$ is an integer then $\vec x_0\in \mathbb U_d^n$ and thus the global maximum  $f(\vec x_0)$  is attainable and in this case $f(\vec x_0)=C_{nd}$.
If $\lambda$ is not an integer then the maximal value $f(\vec x_0)$ is not attainable by any vector in $\mathbb U_d^n$. However its maximum must be attained at those vectors near one of those extremal points that have the floors or ceilings of the components of the extremal points. We consider at first those vectors in $U_d^n$ near $\vec x_0$ that have a number $m$ of $x_+=\left\lceil{\lambda}\right\rceil\theta$ and a number $n-m$ of $x_-=\lfloor\lambda\rfloor\theta$ as components with $n\ge m\ge 0$. On these vectors $f(\vec x)$ assumes values
$
F_m=m\cos x_++(n-m)\cos x_--\cos(m\theta+nx_-).
$
Let $\Delta_m=(F_{m+1}-F_{m})/(2\sin\frac\theta2)$ and we have
\begin{eqnarray}
\Delta_m=\sin \frac{2nx_-+(2m+1)\theta}2-\sin\frac{x_-+x_+}2
\end{eqnarray}
Since $\lambda$ is not an integer we have
 $0\le (x_-+x_+)/2\le \pi/2$. In the case of $d/2>n$ we have $nx_-+m\theta\le 2\pi+x_-$ and then $\Delta_m\ge 0$ if $m<\delta:=d/2-(n+1)\lfloor\lambda\rfloor$ and $\Delta_m\le 0$ if $m>\delta$. As a result $\max F_m=F_\delta=C_{nd}$.
If $n\ge d/2$ then $x_-=0$ and in this case $\Delta_m\ge 0$ if $m^\prime_u\le m\le m_u$ with
$m_u^\prime=ud$ or $m_u=(u+1/2)d-1$ and $\Delta_m< 0$ otherwise for $u\ge 0$ being integer. Thus the maximum value of $F_m$ must be taken
on $m_u^\prime$ or $m_u$ and obviously $F_{m_0}\ge F_{m_u}$ and $F_{m_0}\ge F_{m^\prime_u}$ for all $u$. As a result we have $\max F_m=F_{m_0}=C_{nd}$. Since $C_{nd}\ge (n+1)\cos\lceil \lambda\rceil\theta\ge f(\vec x_l)$ for all $l\ge 1$ we see that $C_{nd}$ is the global maximum of $f$.\hfill $\sharp$

\end{document}